\documentstyle[12pt,epsf]{article}
\newlength{\pubnumber} 

\catcode`\@=11
\@addtoreset{equation}{section}

\def\section{\@startsection{section}{1}{\z@}{3.5ex plus 1ex minus .2ex}
 {2.3ex plus .2ex}{\large\bf}}
\def\subsection{\@startsection{subsection}{2}{\z@}{2.3ex plus .2ex}
 {2.3ex plus .2ex}{\bf}}

\begin{document}

\begin{titlepage}
\samepage{
\setcounter{page}{1}
\rightline{UFIFT--HEP--97--28, UPR-0773-T}
\rightline{\tt hep-ph/9711339}
\rightline{November 1997}
\vfill
\begin{center}
 {\Large \bf  On the Anomalous $U(1)$ in Free Fermionic Superstring Models\\}
\vfill
\vfill {\large
	Gerald B. Cleaver$^{1}$\footnote{gcleaver@langacker.hep.upenn.edu}
        and
        Alon E. Faraggi$^{2}$\footnote{faraggi@phys.ufl.edu}}\\
\vspace{.12in}
{\it $^{1}$ Department of Physics and Astronomy\\
	    University of Pennsylvania\\
	    Philadelphia, PA 19104, USA\\}
\vspace{.06in}
{\it$^{2}$  Institute for Fundamental Theory, \\
            Department of Physics, \\
            University of Florida, \\
            Gainesville, FL 32611, USA\\}
\vspace{.025in}
\end{center}
\vfill
\begin{abstract}
The realistic free fermionic models have had an intriguing
success in explaining different properties of
the observed particle spectrum. In this paper
we discuss in some detail the anomalous $U(1)$ symmetry
which exists in these models.
We study the properties of the anomalous $U(1)$
in both the more realistic NAHE--based free fermionic models and those
in a general NAHE class. Appearance of an anomalous $U(1)$
in the more realistic NAHE models
is shown to be an effect of reduction of world--sheet
supersymmetry from $(2,2)$ to $(2,0)$. We show, however, that in more
general $(2,1)$ and $(2,0)$ models, all $U(1)$ can remain anomaly--free
under certain conditions. Several phenomenological issues related
to the anomalous $U(1)$ are discussed. In particular, we note
that in some examples the anomalous $U(1)$ arises from the breaking
$E_6\rightarrow SO(10)\times U(1)_A$, resulting in $U(1)_A$ being
family universal.
\end{abstract}
\smallskip}
\end{titlepage}

\setcounter{footnote}{0}

% ========================= DEFINITIONS ===================================
\def\beq{\begin{equation}}
\def\eeq{\end{equation}}
\def\beqn{\begin{eqnarray}}
\def\eeqn{\end{eqnarray}}

\def\no{\noindent }
\def\nolabel{\nonumber }
\def\ie{{\it i.e.}}
\def\eg{{\it e.g.}}
\def\half{{\textstyle{1\over 2}}}
\def\third{{\textstyle {1\over3}}}
\def\quarter{{\textstyle {1\over4}}}
\def\sixth{{\textstyle {1\over6}}}
\def\m{{\tt -}}
\def\p{{\tt +}}

\def\Tr{{\rm Tr}\, }
\def\tr{{\rm tr}\, }

\def\slash#1{#1\hskip-6pt/\hskip6pt}
\def\slk{\slash{k}}
\def\GeV{\,{\rm GeV}}
\def\TeV{\,{\rm TeV}}
\def\y{\,{\rm y}}
\def\SM{Standard--Model }
\def\SUSY{supersymmetry }
\def\SSSM{supersymmetric standard model}
\def\vev#1{\left\langle #1\right\rangle}
\def\l{\langle}
\def\r{\rangle}
\def\o#1{\frac{1}{#1}}

\def\Htw{{\tilde H}}
\def\chibar{{\overline{\chi}}}
\def\qbar{{\overline{q}}}
\def\ibar{{\overline{\imath}}}
\def\jbar{{\overline{\jmath}}}
\def\Hbar{{\overline{H}}}
\def\Qbar{{\overline{Q}}}
\def\abar{{\overline{a}}}
\def\alphabar{{\overline{\alpha}}}
\def\betabar{{\overline{\beta}}}
\def\tautwo{{ \tau_2 }}
\def\thetatwo{{ \vartheta_2 }}
\def\thetathree{{ \vartheta_3 }}
\def\thetafour{{ \vartheta_4 }}
\def\ttwo{{\vartheta_2}}
\def\tthree{{\vartheta_3}}
\def\tfour{{\vartheta_4}}
\def\ti{{\vartheta_i}}
\def\tj{{\vartheta_j}}
\def\tk{{\vartheta_k}}
\def\calF{{\cal F}}
\def\smallmatrix#1#2#3#4{{ {{#1}~{#2}\choose{#3}~{#4}} }}
\def\ab{{\alpha\beta}}
\def\Minv{{ (M^{-1}_\ab)_{ij} }}
\def\bone{{\bf 1}}
\def\ii{{(i)}}
\def\V{{\bf V}}
\def\N{{\bf N}}

% for basis vectors:
\def\b{{\bf b}}
\def\S{{\bf S}}
\def\X{{\bf X}}
\def\I{{\bf I}}
\def\mb{{\mathbf b}}
\def\mS{{\mathbf S}}
\def\mX{{\mathbf X}}
\def\mI{{\mathbf I}}
\def\balpha{{\mathbf \alpha}}
\def\bbeta{{\mathbf \beta}}
\def\bgamma{{\mathbf \gamma}}
\def\bxi{{\mathbf \xi}}

\def\t#1#2{{ \Theta\left\lbrack \matrix{ {#1}\cr {#2}\cr }\right\rbrack }}
\def\C#1#2{{ C\left\lbrack \matrix{ {#1}\cr {#2}\cr }\right\rbrack }}
\def\tp#1#2{{ \Theta'\left\lbrack \matrix{ {#1}\cr {#2}\cr }\right\rbrack }}
\def\tpp#1#2{{ \Theta''\left\lbrack \matrix{ {#1}\cr {#2}\cr }\right\rbrack }}
\def\l{\langle}
\def\r{\rangle}

%================== BLACKBOARD BOLD CHARACTERS ==============================

\def\inbar{\,\vrule height1.5ex width.4pt depth0pt}

\def\IC{\relax\hbox{$\inbar\kern-.3em{\rm C}$}}
\def\IQ{\relax\hbox{$\inbar\kern-.3em{\rm Q}$}}
\def\IR{\relax{\rm I\kern-.18em R}}
 \font\cmss=cmss10 \font\cmsss=cmss10 at 7pt
\def\IZ{\relax\ifmmode\mathchoice
 {\hbox{\cmss Z\kern-.4em Z}}{\hbox{\cmss Z\kern-.4em Z}}
 {\lower.9pt\hbox{\cmsss Z\kern-.4em Z}}
 {\lower1.2pt\hbox{\cmsss Z\kern-.4em Z}}\else{\cmss Z\kern-.4em Z}\fi}

%========================================================================
%          MACROS FOR REFERENCES
%========================================================================
\def\AEF{A.E. Faraggi}
\def\NPB#1#2#3{{\it Nucl.\ Phys.}\/ {\bf B#1} (19#2) #3}
\def\PLB#1#2#3{{\it Phys.\ Lett.}\/ {\bf B#1} (19#2) #3}
\def\PRD#1#2#3{{\it Phys.\ Rev.}\/ {\bf D#1} (19#2) #3}
\def\PRL#1#2#3{{\it Phys.\ Rev.\ Lett.}\/ {\bf #1} (19#2) #3}
\def\PRT#1#2#3{{\it Phys.\ Rep.}\/ {\bf#1} (19#2) #3}
\def\MODA#1#2#3{{\it Mod.\ Phys.\ Lett.}\/ {\bf A#1} (19#2) #3}
\def\IJMP#1#2#3{{\it Int.\ J.\ Mod.\ Phys.}\/ {\bf A#1} (19#2) #3}
\def\nuvc#1#2#3{{\it Nuovo Cimento}\/ {\bf #1A} (#2) #3}
\def\RPP#1#2#3{{\it Rept.\ Prog.\ Phys.}\/ {\bf #1} (19#2) #3}
\def\etal{{\it et al\/}}

%==============================================================================
\hyphenation{su-per-sym-met-ric non-su-per-sym-met-ric}
\hyphenation{space-time-super-sym-met-ric}
\hyphenation{mod-u-lar mod-u-lar--in-var-i-ant}
%==============================================================================

%============================== SECTION 1 ============================

\setcounter{footnote}{0}
\section{Introduction}
The more realistic superstring models in the free fermionic
class have had intriguing success in providing plausible
explanations for various properties of the Standard Model
\cite{ffmreviews}.
A few of these include: the natural emergence of three
generations; the origin of the heavy top quark mass;
the qualitative pattern of the fermion mass spectrum;
the stability of the proton; and more \cite{ffmreviews}.

A general property of these models,
which is also shared by many other superstring vacua, is the
existence of an ``anomalous" $U(1)$.
The presence of an Abelian ``anomalous'' symmetry in superstring
derived models yields many desirable phenomenological implications
from the point of view of the effective low energy field theory.
The list of a few of those possibilities include:
Generation of a Fayet--Iliopoulos $D$--term \cite{u1a};
Breaking and rank reduction of the four dimensional gauge group \cite{u1bgg};
Generation of the fermion mass hierarchy \cite{u1sfmh,u1ffmh,ramond};
Predicting neutrino mixing patterns \cite{u1pnmp};
Implication of the relation between anomaly cancelation
and gauge coupling normalization on the weak mixing angle \cite{u1weakangle};
Anomalous $U(1)$ as the trigger of
supersymmetry breaking \cite{u1ssb,u1fsb,fp2};
Cosmological implications of the anomalous $U(1)$ such as
implications for inflation and cosmic strings \cite{u1i};
Anomalous $U(1)$ in connection with the strong CP problem \cite{u1cp}; and
Non--perturbative considerations of the anomalous $U(1)$ \cite{u1np}.

These phenomenological implications make evident the need
to determine what we may actually learn from the anomalous
$U(1)$ symmetries appearing in realistic superstring models.
In this paper we undertake this task by examining
the anomalous $U(1)$ in the NAHE--based \cite{nahe,slm,revamp}
free fermionic models.
We first discuss the origin of the anomalous $U(1)$ and
classify the different sources that may contribute to the
anomaly. Further, we discuss general constraints on a $U(1)$
becoming anomalous and properties of a model that prevent the
appearance of the anomaly.
We then proceed to examine the anomalous
$U(1)$ in specific models and discuss some of the
phenomenological implications on the final gauge group
and on supersymmetry breaking.
An important property of the anomalous $U(1)$ in some of the
free fermionic models is the fact that it is family universal.
In the NAHE--based models, the origin of this universality arises
due to the cyclic permutation symmetry that characterizes the $Z_2\times Z_2$
orbifold, which underlies these realistic free fermionic models.
The contribution to the anomalous $U(1)$
from sectors belonging to the $Z_2\times Z_2$ orbifold
is family universal. This property is often destroyed
when additional basis vectors, beyond those that correspond
to the $Z_2\times Z_2$ orbifold, are added to obtain the full
set of basis vectors necessary for these three generation models.
However, in some models the anomalous $U(1)$ universality is
preserved and may serve as an additional phenomenological
criteria in the classification of this class of models.

\setcounter{footnote}{0}
\section{Realistic free fermionic models}

In this section we recapitulate the main structure of
the free fermionic models. We discuss here the main
features which are important for understanding
the origin or the anomalous $U(1)$ and for
eventually relating the phenomenological implications
of the anomalous $U(1)$ in these models to the
basic structures underlying the models.

In the free fermionic formulation of the heterotic string \cite{heterotic}
all the degrees of freedom needed to cancel the conformal anomaly
are represented in terms of internal free fermions propagating
on the string world--sheet.
In four dimensions, this requires 20 left--moving
and 44 right--moving real world--sheet fermions.
Equivalently, some real fermions may be paired to
form complex fermions.
Under parallel transport around a noncontractible loop,
the fermionic states pick up a phase.
Specification of the phases for all world--sheet fermions
around all noncontractible loops contributes
to the spin structure of the model.
Such spin structures are usually given in the form
of boundary condition ``vectors'', with each element
of the vector specifying the phase of a corresponding
world--sheet fermion.
The possible spin structures are constrained
by string consistency requirements
(e.g. modular invariance). A model is constructed by
choosing a set of boundary condition basis vectors,
which satisfies the modular invariance constraints.
The basis vectors, $\b_k$, span a finite
additive group $\Xi=\sum_k{{n_k}{\b_k}}$
where $n_k=0,\cdots,{{N_{z_k}}-1}$.
The physical massless states in the Hilbert space of a given sector
$\balpha\in{\Xi}$, are obtained by acting on the vacuum with
bosonic and fermionic operators and by
applying the generalized GSO projections. The $U(1)$
charges, $Q(f)$, with respect to the unbroken Cartan
generators of the four
dimensional gauge group, which are in
one--to--one correspondence with the $U(1)$
currents ${f^*}f$ for each complex fermion $f$, are given by:
\beq
{Q(f) = {1\over 2}\balpha(f) + F(f)},
\eeq
where $\balpha(f)$ is the boundary condition of the world--sheet fermion $f$
in the sector $\balpha$, and
$F_\balpha(f)$ is a fermion number operator counting each mode of
$f$ once (and if $f$ is complex, $f^*$ minus once).
For periodic fermions, $\balpha(f)=1$ and
the vacuum is a spinor in order to represent the Clifford
algebra of the corresponding zero modes. For each
periodic complex fermion $f$
there are two degenerate vacua ${\vert +\rangle},{\vert -\rangle}$ ,
annihilated by the zero modes $f_0$ and
${{f_0}^*}$ and with fermion numbers  $F(f)=0,-1$, respectively.

\subsection{The NAHE set}\label{nahesec}
The boundary condition basis vectors which generate the realistic
free fermionic models under discussion
are, in general, divided into two major subsets.
The first subset consists of the NAHE set \cite{nahe,slm,revamp}, which is a
set
of five boundary condition basis vectors denoted $\{{\bf
1},\S,\b_1,\b_2,\b_3\}$.
With `0' indicating Neveu--Schwarz boundary conditions
and `1' indicating Ramond boundary conditions, these vectors are as follows:
\beqn
 &&\begin{tabular}{c|c|ccc|c|ccc|c}
 ~ & $\psi^\mu$ & $\chi^{12}$ & $\chi^{34}$ & $\chi^{56}$ &
        $\overline{\psi}^{1,...,5} $ &
        $\overline{\eta}^1 $&
        $\overline{\eta}^2 $&
        $\overline{\eta}^3 $&
        $\overline{\phi}^{1,...,8} $ \\
\hline
\hline
      {\bf 1} &  1 & 1&1&1 & 1,...,1 & 1 & 1 & 1 & 1,...,1 \\
          $\S$ &  1 & 1&1&1 & 0,...,0 & 0 & 0 & 0 & 0,...,0 \\
\hline
  $ \b_1$ &  1 & 1&0&0 & 1,...,1 & 1 & 0 & 0 & 0,...,0 \\
  $ \b_2$ &  1 & 0&1&0 & 1,...,1 & 0 & 1 & 0 & 0,...,0 \\
  $ \b_3$ &  1 & 0&0&1 & 1,...,1 & 0 & 0 & 1 & 0,...,0 \\
\end{tabular}
   \nonumber\\
   ~  &&  ~ \nonumber\\
   ~  &&  ~ \nonumber\\
     &&\begin{tabular}{c|cc|cc|cc}
 ~&      $y^{3,...,6}$  &
        $\overline{y}^{3,...,6}$  &
        $y^{1,2},\omega^{5,6}$  &
        $\overline{y}^{1,2},\overline{\omega}^{5,6}$  &
        $\omega^{1,...,4}$  &
        $\overline{\omega}^{1,...,4}$   \\
\hline
\hline
    {\bf 1} & 1,...,1 & 1,...,1 & 1,...,1 & 1,...,1 & 1,...,1 & 1,...,1 \\
    $\S$     & 0,...,0 & 0,...,0 & 0,...,0 & 0,...,0 & 0,...,0 & 0,...,0 \\
\hline
$ \b_1$ & 1,...,1 & 1,...,1 & 0,...,0 & 0,...,0 & 0,...,0 & 0,...,0 \\
$ \b_2$ & 0,...,0 & 0,...,0 & 1,...,1 & 1,...,1 & 0,...,0 & 0,...,0 \\
$ \b_3$ & 0,...,0 & 0,...,0 & 0,...,0 & 0,...,0 & 1,...,1 & 1,...,1 \\
\end{tabular}
\label{nahe}
\eeqn
with the following
choice of phases which define how the generalized GSO projections are to
be performed in each sector of the theory:
\beq
      C\left( \matrix{\b_i\cr \b_j\cr}\right)~=~
      C\left( \matrix{\b_i\cr \S\cr}\right) ~=~
      C\left( \matrix{\bone \cr \bone \cr}\right) ~= ~ -1~.
\label{nahephases}
\eeq
The remaining projection phases can be determined from those above through
the self--consistency constraints.
The precise rules governing the choices of such vectors and phases, as well
as the procedures for generating the corresponding space--time particle
spectrum, are given in Refs.~\cite{fff}.

The basis vector $\S$ generates the space--time supersymmetry.
The set of basis vectors $\{{\bf 1},\S\}$ generates a model
with $N=4$ space--time supersymmetry and $SO(44)$ gauge group
in the right--moving sector. Imposing the GSO projections
of the basis vectors $\b_1$, $\b_2$ and $\b_3$ reduces the $N=4$
supersymmetry to $N=1$ and breaks the gauge group to
$SO(10)\times SO(6)^3\times E_8$.
The space--time vector bosons that generate the gauge group
arise from the Neveu--Schwarz sector and from the sector
$\I \equiv {\bf 1}+\b_1+\b_2+\b_3$.
The Neveu--Schwarz sector produces the generators of
$SO(10)\times SO(6)^3\times SO(16)$, while
the sector ${\bf 1}+\b_1+\b_2+\b_3$
produces the spinorial {\bf 128} of $SO(16)$ and completes the hidden
gauge group to $E_8$.
The three basis vectors $\b_1$, $\b_2$ and $\b_3$ correspond to the three
twisted sectors of the $Z_2\times Z_2$ orbifold. Each one of these sectors
produces 16 multiplets in the ${\bf 16}$ representation of $SO(10)$.

While the realistic free fermionic models have $(2,0)$ world--sheet
supersymmetry \cite{wss}, the NAHE set by itself
produces a model with $(2,1)$ world--sheet supersymmetry.
The NAHE set belongs to a small class of heterotic string models
that correspond to type--II models under the $H$--map \cite{bs}.
Consistent type--II strings have at least $(1,1)$ world--sheet
supersymmetry and at least $(2,1)$ world--sheet supersymmetry
if they also possess space--time supersymmetry.
World--sheet supersymmetry is preserved under the $H$--map.
Relatedly, the $H$--map and the associated $(2,1)$ or $(2,2)$
world--sheet supersymmetry reveal that the
twelve internal left-- and right--moving fermions,
$\{y,\omega\}$ and $\{\bar{y},\bar{\omega}\}$,
correspond to the internal six--dimensional compactified torus
in the bosonic formulation, while
$\{\chi\}$ and $\{\bar\eta\}$ are their respective
left-- and right--moving complexified world--sheet
superpartners.
The boundary conditions of all of these fields are unaffected by the
$H$--map. If a heterotic string has at least
an $SO(10)\times E_8$ gauge group (and $N=2$ left--moving world--sheet
supersymmetry), then it is mapped to
a $(2,1)$ type--II sting, whereas
a heterotic string with at least an $E_6\times E_8$ gauge group
is identified with a $(2,2)$ type--II string.
Further, a $E_7\times E_8$ gauge group is mapped to a
$(2,2+4)$ model, while $E_8\times E_8$ corresponds to a model with $(2,2+2+2)$
\cite{wssn}.

The correspondence of the NAHE set with a $Z_2\times Z_2$ orbifold
is illustrated by adding to the NAHE the boundary condition basis
vector $\X$ with periodic boundary conditions for the world--sheet
fermions $\{{\bar\psi}^{1,\cdots,5},{\bar\eta}^{1,2,3}\}$,
and antiperiodic boundary conditions for all others,
\beq
\X=(0,\cdots,0\vert{\underbrace{1,\cdots,1}_{{\psi^{1,\cdots,5}},
{\eta^{1,2,3}}}},0,\cdots,0)~.
\label{vectorx}
\eeq
The choice of generalized GSO projection coefficients is
\begin{equation}
      C\left( \matrix{\X\cr \b_j\cr}\right)~=~
     -C\left( \matrix{\X\cr \S\cr}\right) ~=~
      C\left( \matrix{\X\cr \bone \cr}\right) ~= ~ +1~.
\label{Xphases}
\end{equation}
With the set $\{{\bf 1},\S,\b_1,\b_2,\b_3,\X\}$ the gauge group
is $E_6\times U(1)^2\times SO(4)^3\times E_8$. (Therefore the model
clearly has $(2,2)$ world--sheet supersymmetry).
For our purposes here
it is more convenient to generate the same model with the
basis vectors
\beq
\{{\bf 1},\S,\bxi={\bf 1}+\b_1+\b_2+\b_3,\X,\b_1,\b_2\},
\label{alternativeset}
\eeq
and the generalized GSO projection coefficients
\beq
c\left(\matrix{\b_i\cr
                                    \b_j\cr}\right)=
c\left(\matrix{\b_i\cr
                                    \S\cr}\right)=
c\left(\matrix{\b_i\cr
                                    \bxi,\X\cr}\right)=
-c\left(\matrix{\bxi,\X\cr
                                    \bxi,\X\cr}\right)=
-c\left(\matrix{{\bf 1}\cr
                                    {\bf 1}\cr}\right)=-1.
\label{e6phases}
\eeq
The first four vectors
in the basis $\{{\bf 1},\S,\bxi,\X\}$ generate a model with $N=4$
space--time supersymmetry with an $E_8\times SO(12)\times E_8$ gauge
group. The sector {\bf S} generates  $N=4$ space--time supersymmetry.
The first and second $E_8$ are obtained from the world--sheet
fermionic states $\{{\bar\psi}^{1,\cdots,5},{\bar\eta}^{1,2,3}\}$ and
$\{{\bar\phi}^{1,\cdots,8}\}$, respectively, while $SO(12)$ is obtained
from $\{{\bar y},{\bar\omega}\}^{1,\cdots,6}$.
The Neveu--Schwarz sector
produces the adjoint representations of $SO(16)\times SO(12)\times SO(16)$.
The sectors $\X$ and
$\bxi$ produce the spinorial representation of $SO(16)$ of the
observable and hidden sectors respectively,
and complete the observable and hidden gauge groups to $E_8\times E_8$.

The vectors $\b_1$ and $\b_2$ break the gauge symmetry to
$E_6\times U(1)^2\times SO(4)^3\times E_8$ and $N=4$ to $N=1$
space--time supersymmetry.
We denote the $U(1)$ generators produced by the world--sheet currents
$:{\bar\eta}^{i*}{\bar\eta}^{i}:$ as $U(1)_i$.
The fermionic states $\{\chi^{12},\chi^{34},\chi^{56}\}$ and
$\{{\bar\eta}^1,{\bar\eta}^2,{\bar\eta}^3\}$ give the usual
``standard--embedding", with
$b(\chi^{12},\chi^{34},\chi^{56})=b({\bar\eta}^1,{\bar\eta}^2,{\bar\eta}^3)$.
The $U(1)$ current of the left--moving $N=2$ world--sheet supersymmetry
is given by
\beq
J(z)= \sum_{i=(12),(34),(56)} :\chi^{i*}\chi^{i}: .
\label{lmu1current}
\eeq
The $U(1)_i$ charge combination appearing in the
decomposition of $E_6$ under $SO(10)\times U(1)$ is given by
\beq
U(1)_{E_6}=U(1)_1+U(1)_2+U(1)_3 ,
\label{e6u1}
\eeq
while the charges of the two orthogonal combinations are specified by
\beqn
&& U(1)^\prime=U(1)_1-U(1)_2
\label{orthocoma}\\
&& U(1)^{\prime\prime}=U(1)_1+U(1)_2-2U(1)_3.
\label{orthocom}
\eeqn
The three $SO(4)$ gauge groups are produced by the
right--moving world--sheet
fermionic fields
$\{{\bar y}^{3,\cdots,6}\}$, $\{{\bar y}^1,{\bar y}^2
{\bar\omega}^5,{\bar\omega}^6\}$ and $\{{\bar\omega}^{1,\cdots,4}\}$.

The sectors $(\b_1;\b_1+\X)$, $(\b_2;\b_2+\X)$ and $(\b_3;\b_3+\X)$
each give eight $27$'s of $E_6$. The $(NS;NS+\X)$ sector gives in
addition to the vector bosons and spin two states, three copies of
scalar representations in ${\bf 27}+{\overline{\bf 27}}$ of $E_6$.
The net number of generations is 24 in the {\bf 27} representation of $E_6$.
The same model is constructed in the
orbifold formulation \cite{DHVW} by first constructing the Narain lattice
\cite{NARAIN}
with $SO(12)\times E_8\times E_8$ gauge group $N=4$ supersymmetry.
The gauge is then broken to $E_6\times U(1)^2\times SO(4)^3\times E_8$
after applying the $Z_2\times Z_2$ twisting $SO(12)$ lattice
and the three twisted sector produce 48 fixed points, which correspond
to the 24 generations in the fermionic construction \cite{ztwo}.

\subsection{Beyond the NAHE set}

At the level of the NAHE set the observable gauge group is
$SO(10)\times SO(6)^3$ and the number of generations is 48, sixteen
from each sector $\b_1$, $\b_2$ and $\b_3$. The $SO(6)^3$ symmetries
are horizontal flavor dependent symmetries. To break the gauge
group to the standard model and to reduce the number of generations,
we must add additional
boundary condition basis vectors to the NAHE set. These additional
basis vectors break the $SO(10)$ and the flavor $SO(6)$ gauge symmetries
and in turn reduce the number of generations to three.
In the process, world--sheet supersymmetry is reduced from $(2,1)$
to $(2,0)$ via basis vectors like $\bgamma$ shown
below. $(2,1)$ is broken because the $\{\eta,y,\omega\}$ components
of gamma do not correspond to a symmetry (mod change of sign) of the
right--moving supercurrent
\beqn
\overline{T}_{3/2} &=& \overline{T}^{+1}_{3/2} + \overline{T}^{-1}_{3/2}
\label{t32pm}\\
             &=& \phantom{+}\sum_{i=1}^3
                 \frac{1}{\sqrt{2}}(\bar{\eta}^i +
                 {\bar\eta}^{i\ast})
                  {\bar{y}}^{2i-1}{\bar{\omega}}^{2i-1}
   + \sum_{i=1}^3
    \frac{-i}{\sqrt{2}}({\bar\eta}^i - {\bar{\eta}}^{i\ast})
                  {\bar{y}}^{2i}   {\bar{\omega}}^{2i},
\label{t32}
\eeqn
within the right--moving $N=1$ world--sheet supersymmetry.
$\overline{T}^{+1}_{3/2}$ and $\overline{T}^{-1}_{3/2}$ are
the two supercurrents within $N=2$.
Note, however,
that $2 \bgamma$ is consistent with $N=1$.

The breaking of the gauge group and the
reduction to three generations are done simultaneously.
In fact the reduction to three generations is correlated
with the breaking of the flavor $SO(6)^3$ symmetries to a
product of horizontal $U(1)$ symmetries. The appealing
property of these NAHE--based free fermionic models is that
the emergence of three generations is correlated with
the underlying $Z_2\times Z_2$ orbifold structure.
While three generations can appear from other structures \cite{chl},
here each generation is obtained from one of the twisted sectors
of the $Z_2\times Z_2$ orbifold. Table (\ref{m278}) is an example of
a choice of additional boundary condition basis vectors which
produce a three generation model with $SU(3)\times SU(2)\times U(1)^2$
gauge group. The complete spectrum and charges under the four dimensional
gauge group are given in ref. \cite{eu}.
\beqn
 &\begin{tabular}{c|c|ccc|c|ccc|c}
 ~ & $\psi^\mu$ & $\chi^{12}$ & $\chi^{34}$ & $\chi^{56}$ &
        $\bar{\psi}^{1,...,5} $ &
        $\bar{\eta}^1 $&
        $\bar{\eta}^2 $&
        $\bar{\eta}^3 $&
        $\bar{\phi}^{1,...,8} $\\
\hline
\hline
  ${\balpha}$  &  0 & 0&0&0 & 1~1~1~0~0 & 0 & 0 & 0 & 1~1~1~1~0~0~0~0 \\
  ${\bbeta}$   &  0 & 0&0&0 & 1~1~1~0~0 & 0 & 0 & 0 & 1~1~1~1~0~0~0~0 \\
  ${\bgamma}$  &  0 & 0&0&0 &
		${1\over2}$~${1\over2}$~${1\over2}$~${1\over2}$~${1\over2}$
	      & ${1\over2}$ & ${1\over2}$ & ${1\over2}$ &
                ${1\over2}$~0~1~1~${1\over2}$~${1\over2}$~${1\over2}$~0 \\
  $2{\bgamma}$ &  0 & 0&0&0 &
		1~1~1~1~1  & 1 & 1  & 1 &
                1~0~0~0~1~1~1~0 \\
\end{tabular}
   \nonumber\\
   ~  &  ~ \nonumber\\
   ~  &  ~ \nonumber\\
     &\begin{tabular}{c|c|c|c}
 ~&   $y^3{y}^6$
      $y^4{\bar y}^4$
      $y^5{\bar y}^5$
      ${\bar y}^3{\bar y}^6$
  &   $y^1{\omega}^5$
      $y^2{\bar y}^2$
      $\omega^6{\bar\omega}^6$
      ${\bar y}^1{\bar\omega}^5$
  &   $\omega^2{\omega}^4$
      $\omega^1{\bar\omega}^1$
      $\omega^3{\bar\omega}^3$
      ${\bar\omega}^2{\bar\omega}^4$ \\
\hline
\hline
$\balpha$ & 1 ~~~ 0 ~~~ 0 ~~~ 0  & 0 ~~~ 0 ~~~ 1 ~~~ 1  & 0 ~~~ 0 ~~~ 1 ~~~ 1
\\
$\bbeta$  & 0 ~~~ 0 ~~~ 1 ~~~ 1  & 1 ~~~ 0 ~~~ 0 ~~~ 0  & 0 ~~~ 1 ~~~ 0 ~~~ 1
\\
$\bgamma$ & 0 ~~~ 1 ~~~ 0 ~~~ 1  & 0 ~~~ 1 ~~~ 0 ~~~ 1  & 1 ~~~ 0 ~~~ 0 ~~~ 0
\\
$2\bgamma$& 0 ~~~ 0 ~~~ 0 ~~~ 0  & 0 ~~~ 0 ~~~ 0 ~~~ 0  & 0 ~~~ 0 ~~~ 0 ~~~ 0
\\
\end{tabular}
\label{m278}
\eeqn
with the choice of GSO projection coefficients:
\beq
c\left(\matrix{\b_j\cr
                                    \balpha,\bbeta,\bgamma\cr}\right)=
-c\left(\matrix{\balpha\cr
                                    {\bf 1}\cr}\right)=
c\left(\matrix{\balpha\cr
                                    \bbeta\cr}\right)=
-c\left(\matrix{\bbeta\cr
                                    {\bf 1}\cr}\right)=
c\left(\matrix{\bgamma\cr
                                    {\bf 1},\balpha\cr}\right)=
-c\left(\matrix{\bgamma\cr
                                    \bbeta\cr}\right)= -1
\label{m278phases}
\eeq (j=1,2,3),
with the others specified by modular invariance
and space--time supersymmetry. We comment that the vector
$2\bgamma$ plays a particularly important role in understanding
the origin of the anomalous $U(1)$ in the three generation
free fermionic models, as will be discussed further below.

\subsection{(2,2) $\rightarrow$ (2,1) $\rightarrow$ (2,0) models}

The vector $\X$ combined with the NAHE set produces a model with
$E_6\times SO(4)^3\times U(1)^2\times E_8$ gauge group and with (2,2)
world--sheet supersymmetry. The realistic free fermionic models
under discussion have, at the level of the NAHE set,
an $SO(10)$ symmetry and only $(2,1)$ world--sheet supersymmetry.
While the $(2,1)$ NAHE symmetry is further
broken to $(2,0)$ by a set of non--NAHE basis vectors,
the underlying $(2,2)$ and $(2,1)$
substructures are why we can identify the internal fermions
$\{y, \omega\vert{\bar y},{\bar\omega}\}$ with the compactified
dimensions in the bosonic formulation,
and the $\{\chi,{\bar\eta}\}$ with their complexified
superpartners.

In these models the basis vector $2\bgamma$ replaces the vector $\X$.
The set $\{{\bf 1},\S,\bxi={\bf 1}+\b_1+\b_2+\b_3,2\bgamma\}$
produces a model with $N=4$ supersymmetry and $SO(16)\times
SO(12)\times SO(16)$ gauge group. Alternatively, we can construct
the same model by using the set $\{{\bf 1},\S,\X,\bxi\}$. From this set
we can construct two $N=4$ models which differ by the discrete
choice of the free phase\footnote{In contrast, changing the
the sign of the corresponding $c({2\bgamma\atop\bxi})$
does not alter the gauge group of
$\{{\bf 1},\S,\bxi,2\bgamma\}$.}
\begin{equation}
c\left(\matrix{\X\cr
                        \bxi\cr}\right)=\pm1\, .
\label{22to21}
\end{equation}
For one phase choice $(+)$ the gauge group
is $E_8\times SO(12)\times E_8$,
which is the standard toroidal compactification.
This is a
$(2,2)$\footnote{More specifically the world--sheet
supersymmetry is $(2,2+2+2)$.} model.
However, for the alternate phase choice $(-)$ the states in the
spinorial representation of $SO(16)$ which make up part of
the adjoint of $E_8$ are removed by the GSO projections.
Thus,
we are left with an $SO(16)\times SO(12)\times SO(16)$ gauge
group and apparently $(2,1)$ world--sheet
supersymmetry.\footnote{Since $E_8$ is broken, the
$H$--map cannot be used to justify the claim for $(2,1)$ world--sheet
supersymmetry.
However, unlike the $U(1)$ current of the right--moving
$N=2$ world--sheet supersymmetry,
the existence of the right--moving supercurrent
$\overline{T}_{3/2}$ should
be independent of choice of phases, depending only on the
boundary conditions of the $\{\bar{\eta},\bar{y},\bar{w}\}^i$.
Models with matching boundary conditions for all
$\{\chi,\bar \eta\}$, $\{y,\bar{y}\}^i$, and $\{w,\bar{w} \}^i$
pairs should retain $(2,1)$ world--sheet supersymmetry
even after the right--moving $N=2$ $U(1)$ current
is broken.}

Applying the orbifold twisting to the
$E_8\times SO(12)\times E_8$ model
breaks one of the $E_8$ to $E_6\times U(1)^2$ and keeps the
$(2,2)$ world--sheet supersymmetry.
Applying the same orbifold twisting to the model with the second
choice of phase reduces  $SO(16)\times SO(12)\times SO(16)$ to
$SO(10)\times U(1)^3\times SO(4)^3\times SO(16)$,
similarly keeping $(2,1)$ symmetry.

Alternatively,
we can start with the basis vectors that produce the $E_6\times U(1)^2$
model and then turn on the GSO projection which projects out
the spinorial ${\bf 16}+{\overline{\bf 16}}$ in the adjoint of $E_6$.
We will end with the same model, as the spectrum is invariant
under the reordering of the GSO projections. Thus, the transition
from the $(2,2)$ world--sheet supersymmetry to the $(2,1)$ world--sheet
supersymmetry can be seen to be a result of a discrete choice
of the free phases. This is an important observation because many
of the useful simple relations that are obtained for $(2,2)$ models
can be used for the realistic free fermionic models.

\section{The origin of the anomalous $U(1)$}

To see how the anomalous $U(1)$ arises in the
NAHE--based realistic free fermionic models, let us
recall the $E_6$ model of section \ref{nahesec}.
This model is generated by the NAHE set basis vectors
plus the vector $\X$ in Eq. (\ref{vectorx}) or alternatively
by the set (\ref{alternativeset}).
Let us now consider the model with the vector $\X$
in Eq. (\ref{alternativeset}) replaced with the vector
$2\bgamma$. In this case we obtain a model with
$SO(10)\times U(1)^3\times SO(4)^3\times SO(16)$
gauge group.
The sectors $\b_j$ still produce the 24 multiplets
in the $\bf 16$ representation of $SO(10)$
with $U(1)_j$ charges $Q_j=1/2$.
In this model the sectors $\b_j+2\bgamma$ now
produce 24 multiplets in the $\bf 16$ vectorial
representation of the hidden $SO(16)$
gauge group, which also carry charges
$Q_i=1/2$, $Q_k=1/2$, $i\ne j\ne k$.
Thus, we observe that in this model
$U_1+U_2+U_3$, which
is the $U(1)$ combination (\ref{e6u1})
embedded in $E_6$ in the $E_6$
model, becomes the anomalous $U(1)$ combination
with the total $\Tr Q_A=-576$. The anomalous $U(1)$
is therefore seen to arise in this model due to
the breaking of the right--moving $N=2$ world--sheet
supersymmetry.

Occurrence of a generic anomalous U(1) can depend on
both boundary conditions and free phases.
An interesting variation of an anomalous
model from the basis set (\ref{alternativeset})
can still occur with the phase choice
$c({\X\atop\bxi})= 1$. While this generates the
$E_8\times SO(12) \times E_8$ gauge group and
$(2,2)$ world--sheet supersymmetry
from the first four basis vectors,
changing the sign of $c({\b_1\atop \bxi})$ and/or
that of $c({\b_2\atop \bxi})$ in (\ref{e6phases})
from $-$ to $+$
produces an anomalous model when basis
vectors $\b_1$ and $\b_2$ are included.
In this case the final gauge group is
$E_6\times U(1)^2\times SO(4)^3\times SO(16)$.
While $U(1)_{E_6}$ defined in (\ref{e6u1})
remains non--anomalous (since it is still embedded
in $E_6$), the orthogonal $U(1)'$ and $U(1)''$ in
(\ref{orthocoma},\ref{orthocom}) become anomalous.

Changing the sign of $c({\b_i\atop \bxi})$
effectively transforms the
$SO(16)$ spinor components of the $E_8$
gauge vectors,
coming from the $\bxi$-sector,
into a massless matter state
(the ${\overline{\bf 128}}$ of $SO(16)$),
thereby breaking $E_8$ to $SO(16)$.
Simultaneously, under sign change of $c({\b_i\atop \bxi})$,
the eight copies of the
{\bf 27} rep of the observable sector $E_6$
(along with some states carrying only $SO(4)^3$ and $U(1)_i$
charges)  originating in $(\b_i,\, \b_i + \X)$
are replaced by eight copies
of a $\bf 16$ rep of $SO(16)_{\rm hid}$.

If only the sign of $c({\b_1\atop \bxi})$ and not that of $c({\b_2\atop \bxi})$
is changed (or vice versa), then the sector
$\b_3 = {\bf 1} + \b_1 + \b_2 + \bxi$
produces exactly eight copies of the $\bf 16$ rep of $SO(16)_{\rm hid}$
that also carry $U(1)'$ and $U(1)''$ charge. If both signs change,
then $\b_3$ yields only $SO(4)^3\times U(1)_i$ reps.
In the former case (with $c({\b_1\atop \bxi})$ transformed),
\beq
\Tr (\eta_1,\eta_2,\eta_3) = (-96,192,-96),
\label{trc1}
\eeq
Here
the anomaly can be placed entirely in
$U(1)_A \equiv U(1)_1 - 2 U(1)_2 + U(1)_3$, also leaving
$U(1)'\equiv U(1)_1 - U(1)_3$ as a good symmetry.
In the latter case,
\beqn
\Tr (\eta_1,\eta_2,\eta_3) &=& (-96,192,-96) + (192,-96,-96)
\label{trc2}\\
                           &=& (96,96,-192),
\label{trc3}
\eeqn
which instead leaves
$U(1)'\equiv U(1)_1 - U(1)_2$ as a good symmetry and
$U(1)_A \equiv U(1)_1 +  U(1)_2 - 2 U(1)_3$
as the anomalous one.
$(2,2)$ world--sheet
supersymmetry appears retained in both cases.

These sign changes in $c({\b_i\atop \bxi})$ produce a trace anomaly
because they destroy the delicate charge cancelation  between
the $\b_{i= 1,2,3}$ sectors in the NAHE set.
Relatedly, destruction of this symmetry under these sign changes,
would most likely make NAHE--based $Z_2\times Z_2$
three generation models extremely
difficult, if not impossible, even after addition of non--NAHE sectors.

\section{Contributions to $U(1)_A$ from beyond the NAHE set}

In all models discussed so far
there appears a correlation between hidden sector
$E_8$ breaking and an unembedded\footnote{Henceforth,
unless it specifically stated otherwise, a $U(1)$
under discussion is assumed {\it not} to be embedded in a
non-Abelian gauge group.} $U(1)$ becoming anomalous.
Although broken $E_8$ models often suggest an anomaly
will be found within a $U(1)$, the NAHE--based
models demonstrate that an anomaly need not necessary appear.
As counter examples we consider the model in \cite{fny} and
that of \cite{eu} already discussed in section 2.2.
In the first case,
removing the last basis vector (the $\bgamma$--like one)
produces a model with
\beq
SO(6)\times SO(4)\times U(1)\times SO(4)\times SO(5)^2 \times
\left[ SO(8)\times SO(8)\right]_{\rm hidden},
\label{fnya}
\eeq
while similarly altering the model in section 2.2 produces
a gauge group
\beq
SO(6)\times SO(4)\times U(1)^2\times [SU(2)_2]^4\times
\left[ SO(16)\right]_{\rm hidden}.
\label{af1a}
\eeq
The single $U(1)$ in the first model and both $U(1)$ in
the second model are non--anomalous even though the
world--sheet supersymmetry in both
models has been reduced from $(2,2)$ to $(2,1)$.

While a broken $E_8$ does not
necessarily imply an anomaly
will appear if there are any $U(1)$'s,
an unbroken level--one $E_8$ does imply that an anomalous $U(1)$
{\it cannot} appear.
This is an example of a more general rule:
\vskip .2truecm

\no Theorem 1. A given $U(1)_i$ in the observable (hidden) sector
is anomaly free if there is a hidden (observable) group $\cal G$
such that all massless $\cal G$--charged states have vanishing charge $Q_i$
\cite{bs,kn}.
\vskip .2truecm

While $E_8$ has been broken in the gauge groups (\ref{fnya}) and
(\ref{af1a}), Thm.~1 explains why
none of the $U(1)$ are anomalous in either case.
Prior to the appearance of a $\bgamma$--sector in each model,
all non--trivial reps of the respective hidden sector gauge groups
do not carry additional $U(1)$ charge.
In fact, without a $\bgamma$--sector
there are no non-trivial massless reps of
$SO(8)^2_{\rm hid}$ in (\ref{fnya}), nor of
$SO(16)_{\rm hid}$ in (\ref{af1a}), in the respective models.
In both cases, adding back the $\bgamma$--like basis vectors
(or even just $2\bgamma$) generates massless states that are both
hidden--sector non--singlets and that carry $U(1)$--charges.

Thm.~1 actually specifies conditions more stringent than is necessary
for anomaly prevention in a  $U(1)$.
The necessary and sufficient condition is stated in:
\vskip .2truecm

\no Theorem 2. A given $U(1)_i$ is anomaly--free if there is a simple gauge
group $\cal G$ in a model such that the trace of $Q_i$ over non--trivial
massless reps of $\cal G$ is zero \cite{bs,kn}.
\vskip .2truecm

A $U(1)$ can remain non--anomalous by reason of Thm.~2,
if, for example, there is a $Z_2$ symmetry that transforms states with
$U(1)$ charge $Q$ into states with charge $-Q$,
resulting in only vector--like pairs of states with regard to that
$U(1)$ \cite{kn}.
Consider for instance model 2 of \cite{chl}).
When basis vectors $V_3$, $V_6$, and $V_7$
(as defined in Table A.2 of \cite{chl}) are removed from this model,
five anomaly--free $U(1)$ appear that
demonstrate this type of anomaly cancelation.
Here the gauge group is $E_6\times U(1)^5 \times SO(22)$.
While there are numerous massless {\bf 22}'s of $SO(22)$ and {\bf 27}'s
of $E_6$ in the model carrying non--zero charge under the five $U(1)$,
a $Z_2$ symmetry exists such that for every ({\bf 22} or {\bf 27}) state
with $U(1)^5$ charge vector $\vec{Q}$ there is a respective
({\bf 22} or {\bf 27}) state with charge vector $-\vec{Q}$.

\vskip .2truecm
\noindent Both Thms. 1 and 2 rest on the fact that
modular invariance of string models relates the traces of an anomalous
$U(1)_A$ over {\it all} gauge groups in the model,
\beq
\frac{1}{K_m}\mathop{\Tr}_{{\cal G}_m}\, T(R)Q_A
      =\o{3}{\Tr} \, Q_A^3
      =     {\Tr} \, Q_B^2 Q_A
      =\o{24}{\Tr}\,  Q_A
      \equiv 8\pi^2 \delta_{\rm GS}
\, ,
\label{gs}
\eeq
(generally known as the universal Green-Schwarz (GS) relation),
where $K_m$ is the level of the gauge group ${\cal G}_m$ and
$2T(R)$ is the index of the representation $R$.
(Both anomalous charge $Q_A$ and the anomaly-free charges $Q_B$
have been rescaled so that $K_A=K_B=1$.)
Since $E_8$ level--one has no massless matter states,
an unbroken level--one $E_8$ in a model automatically
declares all $U(1)$ anomaly--free. This relationship between
traces means that if the $U(1)$ charge trace is zero
over the non--trivial reps of a particular simple gauge group,
the trace is zero over the non--trivial reps of {\it all}
simple gauge groups!

Consideration of Thms.~1.~and 2.~for all $U(1)_i$ in
a model leads to:
\vskip .2truecm

\no Theorem 3. A model is completely free of anomalous $U(1)$
if for each $U(1)_i$, there
is at least one gauge group $\cal G$ for which
(a) all non--trivial massless reps of $\cal G$ do not carry $U(1)_i$ charge,
or
(b) the trace of $Q_i$ over all massless non--trivial reps of $\cal G$ is zero
\cite{bs,kn}.
\vskip .2truecm

This implies that ``$U(1)$--quarantined'' models,
defined as models with at least one hidden (observable) sector simple
gauge group whose non--trivial reps carry no observable (hidden) sector
$U(1)$ charge, are guaranteed to be anomaly free, independent of its
world--sheet supersymmetry class.

It is an interesting question whether
free fermionic ``$U(1)$--quarantined''
$SU(3)_C\times SU(2)_L\times U(1)^n$ or semi--GUT models can be found.
If such models exist, can they yield realistic phenomenology?
Of particular interest would be models of this class with a modified
$\bgamma$ that breaks the observable group $SO(6)\times SO(4)$ to
$SU(3)\times SU(2)\times U(1)^2$ yet does not produce
hidden sector states carrying observable $U(1)$ charges.
NAHE class models of this type can indeed be generated.
For example, instead of adding (\ref{m278}) to the NAHE set
(\ref{nahe}),
consider the variation
\beqn
 &\begin{tabular}{c|c|ccc|c|ccc|c}
 ~ & $\psi^\mu$ & $\chi^{12}$ & $\chi^{34}$ & $\chi^{56}$ &
        $\bar{\psi}^{1,...,5} $ &
        $\bar{\eta}^1 $&
        $\bar{\eta}^2 $&
        $\bar{\eta}^3 $&
        $\bar{\phi}^{1,...,8} $\\
\hline
\hline
  ${\balpha'}$ &  1 & 1&0&0 & 1~1~1~0~0 & 1 & 0 & 0 & 1~1~0~0~0~0~0~0 \\
  ${\bbeta'}$  &  1 & 0&1&0 & 1~1~1~0~0 & 0 & 1 & 0 & 1~1~0~0~0~0~0~0 \\
  ${\bgamma'}$ &  0 & 0&0&0 &
		${1\over2}$~${1\over2}$~${1\over2}$~${1\over2}$~${1\over2}$
	      & ${1\over2}$ & ${1\over2}$ & ${1\over2}$
              & 0~0~0~0~0~0~0~0 \\
\end{tabular}
\label{e7}\\
   ~  &  ~ \nonumber\\
   ~  &  ~ \nonumber\\
     &\begin{tabular}{c|c|c|c}
 ~&   $y^3{\bar y}^3$
      $y^4{\bar y}^4$
      $y^5{\bar y}^5$
      $y^6{\bar y}^6$
  &   $y^1{\bar y}^1$
      $y^2{\bar y}^2$
      $\omega^5{\bar\omega}^6$
      $\omega^6{\bar\omega}^6$
  &   $\omega^2{\omega}^3$
      $\omega^1{\bar\omega}^1$
      $\omega^4{\bar\omega}^4$
      ${\bar\omega}^2{\bar\omega}^3$ \\
\hline
\hline
$\balpha'$& 1 ~~~ 0 ~~~ 0 ~~~ 1  & 0 ~~~ 0 ~~~ 1 ~~~ 0  & 0 ~~~ 0 ~~~ 1 ~~~ 0
\\
$\bbeta'$ & 0 ~~~ 0 ~~~ 0 ~~~ 1  & 0 ~~~ 1 ~~~ 1 ~~~ 0  & 0 ~~~ 1 ~~~ 0 ~~~ 0
\\
$\bgamma'$& 0 ~~~ 1 ~~~ 0 ~~~ 0  & 1 ~~~ 0 ~~~ 0 ~~~ 0  & 0 ~~~ 1 ~~~ 1 ~~~ 0
\\
\end{tabular}
   \nonumber\\
\eeqn
instead, with the choice of GSO projection coefficients:
\beqn
c\left(\matrix{\b_i\cr        \b_j\cr}\right) &=&
c\left(\matrix{\b_i\cr          \S\cr}\right)=
-c\left(\matrix{\S\cr           {\bf 1}\cr}\right)=
-c\left(\matrix{{\bf 1}\cr      {\bf 1}\cr}\right)=
c\left(\matrix{\balpha',\bbeta',\bgamma'\cr  1}\right)=  -1\nolabel\\
-c\left(\matrix{\balpha',\bbeta'\cr \S\cr}\right)&=&
 c\left(\matrix{\bgamma'\cr \S\cr}\right)=
 c\left(\matrix{\balpha'\cr            \b_1\cr}\right)=
 c\left(\matrix{\bbeta'\cr             \b_2\cr}\right)=
 c\left(\matrix{\bgamma'\cr            \b_i\cr}\right)=
-c\left(\matrix{\balpha'\cr            \b_2\cr}\right)= -1\nolabel\\
c\left(\matrix{\bbeta'\cr             \b_1\cr}\right)&=&
c\left(\matrix{\balpha',\bbeta'\cr     \b_3\cr}\right)=
-c\left(\matrix{\bbeta'\cr         \balpha'\cr}\right)=
-c\left(\matrix{\bgamma'\cr         \bbeta'\cr}\right)=
c\left(\matrix{\bgamma'\cr        \balpha'\cr}\right)= 1
\label{e7phases}
\eeqn (j=1,2,3),
with the others specified by modular invariance
and space--time supersymmetry.
As usual, the $\balpha'$ and $\bbeta'$ basis vectors
break $SO(10)$ to $SO(6)\times SO(4)$ and
$\bgamma'$ reduces $SO(6)\times SO(4)$
to $SU(3)_C\times SU(2)_L\times U(1)^2$.
Here, however, $\bgamma'$ does not disturb the hidden
sector $E_8$. Instead only  $\balpha'$ and
$\bbeta'$ affect $E_8$ and break it to $E_7\times SU(2)$.
The final observable gauge group is not
$SU(3)_C\times SU(2)_L\times U(1)^n$ though, because
$2\bgamma'$ both enlarges
$SU(3)_C$ to $SU(4)_C$ and adds an extra $SU(2)$ factor.
The complete gauge group is thus,
\beq
SU(4)_C\times SU(2)_L\times SU(2)\times U(1)^4
\times [ E_7 \times SU(2) ]_{\rm hidden}\, .
\label{su4su2c}
\eeq

Unlike $E_8$, $E_7$ has a massless rep at level 1,
the $\bf 56$. Here the only such $E_7$ state is also
a doublet under $SU(2)_{\rm hid}$.
This state receives contributions of
$\frac{3}{4}$ and $\frac{1}{4}$ to its conformal dimension
from these $E_7$ and $SU(2)$ reps,
respectively, making the state exactly massless.
Hence, there is no room for it to
carry a charge under any of the observable sector $U(1)_i$.
Thus, all four of the $U(1)_i$ are anomaly--free by Thm.~1!
The model is chiral with a net 6 = 7 - 1 quark--doublet
generations in $(4,2,1)$ and $(\bar{4},2,1)$ reps of
$SU(4)_C\times SU(2)_L\times SU(2)$.

Additional
basis vectors must be added to this model
to break $SU(4)_C$ to $SU(3)_C$. As long as space-time
supersymmetry is kept, variations of the
phase coefficients from those given in (\ref{e7phases})
either do not alter the gauge group in (\ref{su4su2c})
or only break $E_7$ to $SO(12)\times SU(2)$.
In the latter event, a $U(1)$ anomaly appears simultaneously
with four hidden sector {\bf 12} reps of $SO(12)$ that carry observable
sector $U(1)_i$ charges.

As we have see in section 3, an important consequence of the NAHE set
basis vectors is that they
result in family universality of the anomalous $U(1)$ for a
a suitable choice of the GSO projection coefficients.
This is an outcome of the cyclic permutation symmetry which
characterizes the $Z_2\times Z_2$ orbifold compactification.
However, the set of basis vectors that define the
three generation free fermionic models contains three
additional basis vectors beyond the NAHE set, denoted
typically by $\{\balpha,\bbeta,\bgamma\}$. The immediate question
is whether the additional basis vectors preserve the
family universality of the anomalous $U(1)$.
The answer, in general, is negative. The universality of the
anomalous $U(1)$ may be spoiled from several sources. One
is that the GSO projections of the additional basis vectors
may project part (or all) of the {\bf 16} of $SO(10)$ from some of the sectors
$\b_j$. Thereby the number of states from the given sector $\b_j$
will be reduced and the universality of the anomalous $U(1)$ is
spoiled. However, these cases in general will not lead to three
generation models and therefore are not of great interest.

Another way by which the universality of the anomalous
$U(1)$ can be spoiled is due to the contribution
of the massless states from the sectors which arise
from the additional basis vectors. It is not apparent
how to extract general patterns from these additional
contributions. For our purposes here
we consider, as an example, the model of ref. \cite{fny}.
In this model
the contribution from the NAHE set basis vectors plus
the vector $2\bgamma$ (in the notation of ref. \cite{fny}
this would be the sector $2\bbeta$) contribute universally
to the three $U(1)$'s, $U_1$, $U_2$ and $U_3$, which arise
from the observable $E_8$. Their combined contribution to these
three $U(1)$'s is $(24,24,-24)$. The contribution from the
other sectors (see ref. \cite{fny} for details of the spectrum)
spoils the family universality of the anomalous $U(1)$
to give $\Tr U_1=-24$, $\Tr U_2= -30$, $\Tr U_3=18$,
$\Tr U_5=6$, $\Tr U_6=6$ and  $\Tr U_8=12$.
The rotated $U(1)$ symmetry containing the entire anomaly
is given by (\ref{anomau1infny}).
Examination of several of the
other models which appear in the
literature \cite{fny,slm,chl,gc97,cceel2}
show that the anomalous $U(1)$, in the generic case, is indeed
not universal. It is therefore quite interesting that there
exist some examples in which the anomalous $U(1)$
is family universal. The implications of the anomalous
$U(1)$ in connection with supersymmetry breaking and
flavor changing neutral currents will be reported in ref. \cite{fp2}.

\section{Phenomenological considerations}

In this section we discuss some of the phenomenological
implications related to the existence of
an anomalous $U(1)$ in superstring models.
The existence of an anomalous $U(1)$ in superstring models
has important implications on the phenomenological
properties of the models.
The anomalous $U(1)$ generates a Fayet--Iliopoulos
D--term which breaks supersymmetry at the Planck scale.
Supersymmetry is restored by assigning some non--vanishing
VEVs to a set of Standard Model singlets in the
massless string spectrum which are charged
under the anomalous $U(1)$ and cancel the D--term.
As in general these fields are also charged under
the anomaly free $U(1)$ combinations, cancelation
of the anomaly free D--term equations results in a set
of non--trivial constraints on the allowed pattern of
non--vanishing VEVs. In addition, F--flatness constraints
impose that the superpotential and all of its derivatives
vanish as well. To insure a supersymmetric vacuum, a chosen
pattern of VEVs must then satisfy all the D--term and F--term
flatness constraints. The mechanism by which the anomalous
$U(1)$ D--term equations is canceled is often referred to in the
literature as the Dine--Seiberg--Witten (DSW) mechanism \cite{u1a}.
The scale associated with this breaking is therefore often
referred to as the DSW scale.

As we noted in the introduction,
some of the phenomenological implications of the anomalous $U(1)$
symmetry, and possible patterns of VEVs,
have been investigated in specific models.
Most notable among those is perhaps the generation
of hierarchical fermion masses. Since the allowed fermion mass
terms are restricted by the Abelian and discrete symmetries
in the string models, the fermion mass terms for the lighter
generations are obtained from nonrenormalizable terms
which are suppressed by inverse powers of the effective
string scale. The relevant nonrenormalizable terms also
contain fields which obtain non--vanishing VEV
in the application of the DSW mechanism. Then some
of the nonrenormalizable terms become effective renormalizable
operators. Successive orders are suppressed by the ratio
of these VEVs to the effective string scale. In this manner
a hierarchical fermion mass spectrum can be generated.
In this paper, however, our interest is not in a
possible pattern of VEVs but rather in what we may
learn by studying the $U(1)$ anomalies, which appear in specific
models, and possibly the set of fields and their charges
under the relevant $U(1)$ symmetries.

In general, a number of $U(1)$ symmetries may appear
anomalous in the original basis in which each
$U(1)_i$ current is generated by a single complex world
sheet free fermion. Of the $n$ anomalous $U(1)$'s
we can always form $n-1$ anomaly free combinations
by rotating all the anomaly into a single, unique anomalous $U(1)_A$,
specified by
\beq
U(1)_A \equiv k \sum_i \{\Tr\, Q_i \} U(1)_i,
\label{srta}
\eeq
where $k$ is a normalization coefficient.
Obviously,
all the possible combinations of anomalous $U(1)$ symmetries
in the original basis, which are anomalous,
must be broken at the DSW scale. Then by examining
the combinations of the anomalous $U(1)$'s which remain
anomalous in (\ref{srta})
at the string scale we can learn some
important lessons with regard to the final gauge
group that persist after application of the DSW
anomaly cancelation mechanism.

An example of this is provided in the model whose
basis vectors and one--loop GSO phases are given in
ref. \cite{fny}. With the choice of GSO projection
coefficients given in ref. \cite{fny},
the anomalous $U(1)$ symmetries are:
${\Tr U_1=-24}$, ${\Tr U_2=-30}$, ${\Tr U_3=18}$,
${\Tr U_5=6}$, ${\Tr U_6=6}$ and  ${\Tr U_8=12}$.
Thus, the entire anomaly can be rotated into
\beq
U_A=-4U_1-5U_2+3U_3+U_5+U_6+2U_8.
\label{anomau1infny}
\eeq
Changing the phase coefficient,
\beq
c\left(\matrix{\b_4\cr
               {\bf 1}\cr}\right)=+1~\rightarrow~
            c\left(\matrix{\b_4\cr
               {\bf 1}\cr}\right)=-1,
\label{pb4tompb4}
\eeq
alters the anomalous $U(1)$'s to: ${\Tr U_C=}-18$, ${\Tr U_L}=12$,
${\Tr U_1=}-18$, ${\Tr U_2=}-24$, ${\Tr U_3=}24$,
${\Tr U_4=}-12$, ${\Tr U_5=}6$, ${\Tr U_6}=6$, ${\Tr U_7}=-6$,
${\Tr U_8=}12$, and ${\Tr U_9}=18$.
The anomalous $U(1)$, thus, transforms into
\beq
U_A=-3U_C+2U_L-3U_1-4U_2+4U_3-2U_4+U_5+U_6-U_7+2U_8+3U_9.
\label{modfny}
\eeq
This modification has an important phenomenological implication.
In the model with the phase modification, Eq. (\ref{pb4tompb4}),
while the weak hypercharge combination,
$U_Y=1/3U_C+1/2U_L$, is anomaly free, the orthogonal
combination, which is embedded in $SO(10)$,
$U_{Z^\prime}=U_C-U_L$ is anomalous. This implies that there exist
models in which this $U_{Z^\prime}$ must be broken near the Planck
scale. Therefore, in such models the universal part of the
observable gauge group, arising from the $SO(10)$ gauge group of the
NAHE set, must be the Standard Model gauge group.
A combination of the flavor dependent $U(1)$'s, however,
may still remain unbroken.

This simple example hinges on a different, more general and
important question. Namely, in specific models given
the initial set of anomalous $U(1)$ symmetries and massless particle
spectrum, what is the final gauge group which is
consistent with the constraints imposed by the requirement
of vanishing F-- and D--flatness constraints, and hence
with the requirement of $N=1$ space--time supersymmetry
at the string scale? For example, in some models it
may be found that the entire four dimensional gauge group
must be broken in order to satisfy the cubic level F-- and
D--flatness constraints. Such models obviously cannot
yield a realistic superstring model, as the Standard Model
gauge group must be broken as well. The task then is to try to
classify such models in terms of the world--sheet boundary
conditions, and possibly by other phenomenological properties
of the models.

The question of the connection between models which allow
specific types of anomalous D--term solutions and other
phenomenological criteria may have important implications.
One of these possible implications is in regard to the
cubic level Yukawa couplings in the superstring derived
standard--like models, which utilize the NAHE set.
The cubic level Yukawa couplings for the quarks and leptons,
in a sector $\b_j$ of a NAHE--based model, are
determined by the boundary conditions in the vector $\bgamma$,
which breaks $SO(10)~\rightarrow~SU(5)\times U(1)$,
according to the following rule \cite{slm}
\begin{eqnarray}
\Delta_j  &=& \vert\bgamma(U(1)_{L_{j+3}})-\bgamma(U(1)_{R_{j+3}})\vert
{\hskip 1cm}(j=1,2,3) \label{udrule}\\
    &=& \left\{ {\begin{array}{ll}
         0 & \rightarrow d_jQ_jh_j+e_jL_jh_j;\\
         1 &\rightarrow u_jQ_j{\bar h}_j+N_jL_j{\bar h}_j,
         \end{array}}
        \right.
\label{udruleb}
\end{eqnarray}
where $\bgamma(U(1)_{R_{j+3}})$, $\bgamma(U(1)_{L_{j+3}})$ are the boundary
conditions of the world--sheet fermionic currents that generate the
$U(1)_{R_{j+3}}$, $U(1)_{L_{j+3}}$ symmetries, respectively.
Thus, we can construct models in which both up and down
type quarks obtain a cubic level mass term.
On the other hand, in models with $\Delta_{1,2,3}=1$ only $+{2\over3}$ charged
type quarks get a cubic level Yukawa coupling.
The distinction between these two types of models
may in fact be correlated with the types of
anomalous $U(1)$ D-term solutions.
To exemplify this issue we consider the model in table
(\ref{model2}).
\beqn
 &\begin{tabular}{c|c|ccc|c|ccc|c}
 ~ & $\psi^\mu$ & $\chi^{12}$ & $\chi^{34}$ & $\chi^{56}$ &
     $\bar{\psi}^{1,...,5}  $ &
     $\bar{\eta}^1 $&
     $\bar{\eta}^2 $&
     $\bar{\eta}^3 $&
     $\bar{\phi}^{1,...,8} $ \\
\hline
\hline
  ${\balpha}$  &  1 & 1&0&0 & 1~1~1~0~0 & 0 & 0 & 0 & 1~1~1~1~0~0~0~0 \\
  ${\bbeta}$   &  1 & 0&1&0 & 1~1~1~0~0 & 0 & 0 & 0 & 1~1~1~1~0~0~0~0 \\
  ${\bgamma}$  &  1 & 0&0&1 &
		${1\over2}$~${1\over2}$~${1\over2}$~${1\over2}$~${1\over2}$
	      & ${1\over2}$ & ${1\over2}$ & ${1\over2}$ &
                ${1\over2}$~0~1~1~${1\over2}$~${1\over2}$~${1\over2}$~1 \\
\end{tabular}
   \nonumber\\
   ~  &  ~ \nonumber\\
   ~  &  ~ \nonumber\\
     &\begin{tabular}{c|c|c|c}
 ~&   $y^3{\bar y}^3$
      $y^4{\bar y}^4$
      $y^5{\bar y}^5$
      ${y}^6{\bar y}^6$
  &   $y^1{\bar y}^1$
      $y^2{\bar y}^2$
      $\omega^5{\bar\omega}^5$
      ${\omega}^6{\bar\omega}^6$
  &   $\omega^2{\omega}^3$
      $\omega^1{\bar\omega}^1$
      $\omega^4{\bar\omega}^4$
      ${\bar\omega}^2{\bar\omega}^3$ \\
\hline
\hline
$\balpha$ & 1 ~~~ 0 ~~~ 0 ~~~ 1  & 0 ~~~ 0 ~~~ 1 ~~~ 0  & 0 ~~~ 0 ~~~ 1 ~~~ 1
\\
$\bbeta$  & 0 ~~~ 0 ~~~ 0 ~~~ 1  & 0 ~~~ 1 ~~~ 1 ~~~ 0  & 0 ~~~ 1 ~~~ 0 ~~~ 1
\\
$\bgamma$ & 1 ~~~ 1 ~~~ 0 ~~~ 0  & 1 ~~~ 1 ~~~ 0 ~~~ 0  & 0 ~~~ 0 ~~~ 0 ~~~ 1
\\
\end{tabular}
\label{model2}
\eeqn
The choice of generalized GSO coefficients is:
\beq
c\left(\matrix{\b_1,\b_3,\balpha,\bbeta,\bgamma\cr
                                    \balpha\cr}\right)=
-c\left(\matrix{\b_2\cr
                                    \balpha\cr}\right)=
 c\left(\matrix{{\bf 1},\b_j,\bgamma\cr
                                    \bbeta\cr}\right)=
-c\left(\matrix{\bgamma\cr
                                    {\bf 1},\b_1,\b_2\cr}\right)=
c\left(\matrix{\bgamma\cr
                                    \b_3\cr}\right)=-1\nonumber
\label{model2c}
\eeq
where $j=1,2,3,$.
The remaining coefficients are specified by modular invariance and
space--time supersymmetry. By the rule (\ref{udrule}) this model produces
down--type Yukawa couplings from the sectors $\b_1$ and $\b_2$
and up--type Yukawa coupling from the sector $\b_3$.
The model contains eight $U(1)$ symmetries, six in the observable
sector and two in the hidden sector. Out of those eight, four are anomaly free
and four are anomalous:
\beq
\Tr \,U_1 = 18\quad \Tr \,U_2 = 30,\quad
\Tr \,U_3 = 24\quad \Tr \,U_4 = 12\, .
\label{eu1}
\eeq
The two $U(1)$'s, $U(1)_L$ and $U(1)_C$, which are embedded in $SO(10)$,
are anomaly free.
Consequently, the weak hypercharge and the orthogonal combination,
$U(1)_{Z^{\prime}}$, are
anomaly free. Likewise, the two $U(1)$'s in the hidden sector are
anomaly free. The anomalous combination is given by:
\beq
U_A=3U_1+5U_2+4U_3+2U_4,\,\,\, {\rm with}\,\,\, \Tr Q_A=318.
\eeq
The three anomaly-free orthogonal combinations are not unique. Different
choices are related by orthogonal transformations. One choice is given by:
\beqn
{U^\prime}_1&=&U_1+U_2-2U_3\\
{U^\prime}_2&=&U_1-U_2+ U_4\\
{U^\prime}_3&=&3U_1-U_2+U_3-4U_4.
\eeqn

The solutions to the D-- and F-- flatness constraints
can be divided into two types. The first utilizes VEVs
of fields only from the Neveu--Schwarz and the
$\zeta=\b_1+\b_2+\balpha+\bbeta$ sectors.
Such solutions are stable to all
orders of nonrenormalizable terms. The second
allows non-vanishing VEVs for matter states from the
sectors $\b_j+2\bgamma$ and for matter states which
break the $U(1)_{Z'}$ which is embedded in $SO(10)$.
In the second type of solutions higher order
nonrenormalizable terms will in general modify
the cubic level F--flat directions.
By examining the massless states from the Neveu--Schwarz
and the $\zeta$ sectors,
it is observed that the number of fields, with independent
charges along the four D constraints is less than four in this model.
The Neveu Schwarz sector produces only one field, $\Phi_{12}$. The sector
$\zeta$ gives $\Phi_{45}$ and $\Phi_3^\pm$, while $\phi_{1,2}$ and
$\phi^\prime_{1,2}$ have the same charges, up to a multiplicative
constant, as $\Phi_{12}$. However, only three of the four fields
have independent charges. The complex conjugate fields can be
used to relax the positive definite restriction but do not add
more degrees of freedom.
Thus the number of constraints is larger than the number of fields which
can be used to solve them.
Adding the states from the sectors $\b_j+2\bgamma$ does not resolve
the problem, since they carry positive charge along the
anomalous $U(1)_A$.
Therefore,
it is found that the number of independent
constraints is larger than the number of flat directions,
implying that
solutions of the first type do not exist in this model.
This example illustrates the connection
between the type of D-- and F--flat solutions, allowed
in specific models, and other phenomenological
characteristics of the models.

The next important phenomenological issue in regard to the anomalous $U(1)$
is supersymmetry breaking. As stated,
the anomalous $U(1)$ generates a Fayet--Iliopoulos
term that break supersymmetry near the Planck scale,
and destabilizes the vacuum. Supersymmetry is restored and the
vacuum is stabilized by giving VEVs to a set
of standard model singlets along F-- and D--flat directions.
If these standard model singlet fields get a mass
term $m$ from nonrenormalizable terms their VEV in the minimum
of the potential may be shifted, resulting in a non--vanishing
$\langle D_A\rangle$ of order $m$. This issue is discussed in detail in
ref.~\cite{fp2}. Here we would like to highlight some
important properties of the anomalous $U(1)$ in this regard.
Note that the $U(1)$ symmetries are in general family dependent,
as they are external to the universal $SO(10)$ subgroup.
Thus, in general the anomalous $U(1)$ is family dependent.
This is a dangerous situation as a non--universal
$\langle D_A\rangle$, even of the order of the electroweak scale,
will give rise to non--degenerate squark masses and therefore
to flavor changing neutral currents in gross excess of the
experimental bounds.

We note, however, that in the free fermionic models the situation
is much improved. This is a consequence of the structure
of the NAHE set of boundary condition basis vectors and the
underlying $Z_2\times Z_2$ orbifold. We first observe that
at the level of NAHE + $2\bgamma$ model, the anomalous
$U(1)$ which is the combination $U_1+U_2+U_3$ is
family universal. This is a consequence of the permutation
symmetry between the NAHE set basis vectors $\b_1$, $\b_2$, and $\b_3$,
which is a reflection of the permutation symmetry between
the three twisted sectors of the $Z_2\times Z_2$ orbifold.
Thus, at this level the universality of the
anomalous $U(1)$ is a generic result of the NAHE set
basis vectors.

The realistic free fermionic models are obtained
by adding several additional basis vectors to the
NAHE set, which break the gauge group and reduce
the number of generations to three,
one from each of the sectors $\b_1$, $\b_2$ and $\b_3$.
As we have seen above, these additional basis vectors
may give rise to an additional massless spectrum which
contributes to the charge trace of the anomalous $U(1)$. Therefore,
the additional basis vectors may, in fact, spoil
the permutation universality property which exist
at the level of the NAHE set, by modifying the
the traces of the individual $U(1)$ currents.
Thus, the anomalous $U(1)$ combination,
in terms of the $U(1)$ world--sheet currents,
may be modified.
In general, therefore, the resulting anomalous
$U(1)$ combination will not remain family universal.
Remarkably, and perhaps not surprisingly, there exist
three generation models which preserve the
permutation universality property of the sectors
$\b_1$, $\b_2$ and $\b_3$, even after the additional
basis vectors, beyond the NAHE set, are included.
In these three generation models, therefore, the anomalous $U(1)$
is family universal. In turn we may envision that the
constraints imposed by FCNC will severely restrict
the allowed anomalous $U(1)$ combinations emerging
from string models, and consequently the allowed
boundary condition basis vectors beyond the NAHE set.
In table \ref{anomalouscharges} we show
the three generation charges under the anomalous
and anomaly free combinations in the model of ref. \cite{eu}.
As can be seen from the table, the charges of
the chiral generations with respect to the anomalous
$U(1)$ are indeed family universal. We also note from
the table that the charges with respect to some
of the other, anomaly free, $U(1)$ combinations
are not family universal. One may then worry that
possible non--vanishing $D$--terms of the anomaly free
combinations may produce non--universal squark
masses and hence FCNC. In ref.~\cite{fp2} this
issue in investigated in detail, where it is shown
that indeed such $D$--terms may arise after
supersymmetry breaking, although it is often found that
the non--universal contributions are suppressed
relative to the universal one.

The structure exhibited by the anomalous $U(1)$ charges
of the chiral generations in the model of ref. \cite{eu}
is a reflection of the underlying $Z_2\times Z_2$
orbifold structure, which in this model is preserved
also when the basis vectors $\{\balpha,\bbeta,\bgamma\}$
are added to the NAHE set basis vectors. This situation
is further exemplified in the model of ref. \cite{top}
where the anomalous, and anomaly free $U(1)$ combinations
are
\beqn
U_A &=& U_1+U_2+U_3\nonumber\\
U_1^\prime &=& U_1-U_2\nonumber\\
U_2^\prime &=& U_1+U_2-2U_3
\label{u1acomb274}
\eeqn
The anomalous $U(1)$ combination in this model is precisely the
combination in the decomposition $E_6~\rightarrow~SO(10)\times U(1)$
that we encountered in considering the transition from
the $(2,2)$ models to the $(2,1)$ models in section 2.
The orthogonal anomaly free combinations are on the other
hand those which are embedded in the $SU(3)$ holonomy group.
In passing, it is intriguing to note that similar form of $U(1)$ combinations
have recently been discussed in connection with the family
mass hierarchy, from a purely phenomenological point of view
\cite{ramond}. Of course, the charges of the chiral generation
under these $U(1)$'s do not necessarily coincide, as those studies
assume a minimal field content. Nevertheless, we may infer that
fermion mass spectrum, together with the tight constraints imposed
by anomaly cancelations, may yield a rather constrained structure.
As we have seen, the free fermionic models, with their characteristic
$Z_2\times Z_2$ orbifold structure, precisely give rise to such a
structure.

\section{Conclusions}

In this paper we showed how the anomalous $U(1)$ symmetry
arises in the more realistic NAHE--based free fermionic models as a
consequence of the transition from the $(2,2)~\rightarrow~(2,0)$ models.
In contrast, we also demonstrated by specific examples
that general $(2,1)$ and $(2,0)$
NAHE--based models containing $U(1)$'s
can be anomaly-free under certain conditions.
General constraints prohibiting appearance of anomalies
were discussed.
In particular, we showed that a model
is guaranteed to be anomaly free
when each of its $U(1)_i$ can be paired with at least one other
gauge group, such that all the non-trivial reps of the latter
group do not additionally carry a $U(1)_i$ charge.
We then presented as an example a semi-GUT model of this type.

We discussed a simple case were the anomalous $U(1)$ is identified with
the $U(1)$ in the decomposition $E_6\rightarrow SO(10)\times U(1)$.
This simple case however has far reaching phenomenological
implications. In three generation models which preserve
some of the initial $Z_2\times Z_2$ orbifold structure,
the $U(1)\subset E_6$ is, or is part of,
the anomalous $U(1)_A$ symmetry.
Consequently, in these cases the anomalous $U(1)$ is
universal, and its contribution to the squark masses
is family independent. The generation mass hierarchy is
still generated from two sources. One is due to the
non--universality of the orthogonal combinations.
The other is due to the multiplicity of Higgs doublets
at the string level. Recent studies have shown that
combinations of anomalous and anomaly free $U(1)$'s,
similar to those that appear in the NAHE-based free fermionic models,
can yield minimal parameterization of the fermion mass spectrum.
These considerations provide further evidence for
the potential relevance of the free fermionic models
in nature.

\section{Acknowledgments}

We would like to thank Jogesh Pati, Pierre Ramond,
Paul Langacker, and Mirjam Cveti\v c for valuable discussions.
This work is supported in part
by DOE Grant No. DE--FG--058ER40272.

%=========================================================================
%======================== REFERENCES =====================================
%=========================================================================

\vfill\eject

\bigskip
\medskip

\bibliographystyle{unsrt}

\vfill\eject
\textwidth=7.5in
\oddsidemargin=-18mm
\topmargin=-5mm
\renewcommand{\baselinestretch}{1.3}
\pagestyle{empty}
\begin{table}
\begin{eqnarray*}
\begin{tabular}{|c|c|c|rrrrrrrr|c|rr|}
\hline
  $F$ & SEC & $SU(3)_C\times SU(2)_L$&$Q_{C}$ & $Q_L$ & $Q_A$ &
   $Q^\prime_1$ & $Q^\prime_2$ & $Q^\prime_3$ & $Q^\prime_4$ &
   $Q^\prime_5$ & $SU(5)_H\times SU(3)_H$ & $Q_{7}$ & $Q_{8}$ \\
\hline
   $L_1$ & $\b_1$      & $(1,2)$ & $-{3\over2}$ & $0$ &
   ${1\over2}$ & ${1\over2}$ & ${1\over2}$ &
				${1\over2}$ & ${1\over2}$ & ${3\over2}$ &
   $(1,1)$ & $0$ & $0$ \\
\hline
   $Q_1$ &            & $(3,2)$ & $ {1\over2}$ & $0$ &
   ${3\over2}$ & ${1\over2}$ & ${1\over2}$ &
				$-{1\over2}$ & $-{1\over2}$ & $-{1\over2}$ &
   $(1,1)$ & $0$ & $0$ \\
\hline
   $d_1$ &            & $({\bar 3},1)$ & $-{1\over2}$ & $-1$ &
   ${3\over2}$ & ${1\over2}$ & ${1\over2}$ &
				$-{1\over2}$ & $-{1\over2}$ & $-{1\over2}$ &
   $(1,1)$ & $0$ & $0$ \\
\hline
   ${N}_1$ &            & $(1,1)$ & ${3\over2}$ & $-1$ &
   ${3\over2}$ & ${1\over2}$ & ${1\over2}$ &
				$-{1\over2}$ & $-{1\over2}$ & $-{1\over2}$ &
   $(1,1)$ & $0$ & $0$ \\
\hline
   $u_1$ &            & $({\bar 3},1)$ & $-{1\over2}$ & $1$ &
   ${1\over2}$ & ${1\over2}$ & ${1\over2}$ &
				${1\over2}$ & ${1\over2}$ & ${3\over2}$ &
   $(1,1)$ & $0$ & $0$ \\
\hline
   ${e}_1$ &            & $(1,1)$ & ${3\over2}$ & $1$ &
   ${1\over2}$ & ${1\over2}$ & ${1\over2}$ &
				${1\over2}$ & ${1\over2}$ & ${3\over2}$ &
   $(1,1)$ & $0$ & $0$ \\
\hline
\hline
   $L_2$ & $\b_2$      & $(1,2)$ & $-{3\over2}$ & $0$ &
   ${1\over2}$ & $-{1\over2}$ & ${1\over2}$ &
				$-{1\over2}$ & ${1\over2}$ & ${3\over2}$ &
   $(1,1)$ & $0$ & $0$ \\
\hline
   $Q_2$ &            & $(3,2)$ & $ {1\over2}$ & $0$ &
   ${3\over2}$ & $-{1\over2}$ & ${1\over2}$ &
				${1\over2}$ & $-{1\over2}$ & $-{1\over2}$ &
   $(1,1)$ & $0$ & $0$ \\
\hline
   $d_2$ &            & $({\bar 3},1)$ & $-{1\over2}$ & $-1$ &
   ${3\over2}$ & $-{1\over2}$ & ${1\over2}$ &
				${1\over2}$ & $-{1\over2}$ & $-{1\over2}$ &
   $(1,1)$ & $0$ & $0$ \\
\hline
   ${N}_2$ &            & $(1,1)$ & ${3\over2}$ & $-1$ &
   ${3\over2}$ & $-{1\over2}$ & ${1\over2}$ &
				${1\over2}$ & $-{1\over2}$ & $-{1\over2}$ &
   $(1,1)$ & $0$ & $0$ \\
\hline
   $u_2$ &            & $({\bar 3},1)$ & $-{1\over2}$ & $1$ &
   ${1\over2}$ & $-{1\over2}$ & ${1\over2}$ &
				$-{1\over2}$ & ${1\over2}$ & ${3\over2}$ &
   $(1,1)$ & $0$ & $0$ \\
\hline
   ${e}_2$ &            & $(1,1)$ & ${3\over2}$ & $1$ &
   ${1\over2}$ & $-{1\over2}$ & ${1\over2}$ &
				$-{1\over2}$ & ${1\over2}$ & ${3\over2}$ &
   $(1,1)$ & $0$ & $0$ \\
\hline
\hline
   $L_3$ & $\b_3$      & $(1,2)$ & $-{3\over2}$ & $0$ &
   ${1\over2}$ & $0$ & $-1$ &
				$0$ & $-1$ & ${3\over2}$ &
   $(1,1)$ & $0$ & $0$ \\
\hline
   $Q_3$ &            & $(3,2)$ & $ {1\over2}$ & $0$ &
   ${3\over2}$ & $0$ & $-1$ &
				$0$ & $1$ & $-{1\over2}$ &
   $(1,1)$ & $0$ & $0$ \\
\hline
   $d_3$ &            & $({\bar 3},1)$ & $-{1\over2}$ & $-1$ &
   ${3\over2}$ & $0$ & $-1$ &
				$0$ & $1$ & $-{1\over2}$ &
   $(1,1)$ & $0$ & $0$ \\
\hline
   ${N}_3$ &            & $(1,1)$ & ${3\over2}$ & $-1$ &
   ${3\over2}$ & $0$ & $-1$ &
				$0$ & $1$ & $-{1\over2}$ &
   $(1,1)$ & $0$ & $0$ \\
\hline
   $u_3$ &            & $({\bar 3},1)$ & $-{1\over2}$ & $1$ &
   ${1\over2}$ & $0$ & $-1$ &
				$0$ & $-1$ & ${3\over2}$ &
   $(1,1)$ & $0$ & $0$ \\
\hline
   ${e}_3$ &            & $(1,1)$ & ${3\over2}$ & $1$ &
   ${1\over2}$ & $0$ & $-1$ &
				$0$ & $-1$ & ${3\over2}$ &
   $(1,1)$ & $0$ & $0$ \\
\hline
\end{tabular}
\label{anomalouscharges}
\end{eqnarray*}
\caption{The three generation charges under the anomalous and
anomaly free $U(1)$ combinations in the model of ref. \cite{eu}.
The anomalous and anomaly free combinations are given by \cite{eu}:}
\end{table}
$~~~~~~~~~{Q}_A       =2Q_1+2Q_2+2Q_3-Q_4-Q_5-Q_6~,~~$

$~~~~~~~~~{Q^\prime}_1=~Q_1-Q_2~,~~$

$~~~~~~~~~{Q^\prime}_2=~Q_1+Q_2-2Q_3~,~~$

$~~~~~~~~~{Q^\prime}_3=~Q_4-Q_5~,~~$

$~~~~~~~~~{Q^\prime}_4=~Q_4+Q_5-2Q_6~,~~$

$~~~~~~~~~{Q^\prime}_5=~Q_1+Q_2+Q_3+2Q_4+2Q_5+2Q_6~.$
\end{document}